\pdfoutput=1
\documentclass[conference]{IEEEtran}
\IEEEoverridecommandlockouts
\usepackage{cite}
\usepackage{amsmath,amsfonts,bm}

\def\eqref#1{equation~\ref{#1}}
\def\1{\bm{1}}

\DeclareMathAlphabet{\mathsfit}{\encodingdefault}{\sfdefault}{m}{sl}
\SetMathAlphabet{\mathsfit}{bold}{\encodingdefault}{\sfdefault}{bx}{n}

\def\emE{{E}}

\def\emG{{G}}

\DeclareMathOperator*{\argmax}{arg\,max}
\DeclareMathOperator*{\argmin}{arg\,min}

\usepackage{booktabs}
\usepackage{mathtools}
\usepackage{algorithmic}
\usepackage{graphicx}
\usepackage{textcomp}
\usepackage{xcolor}
\def\BibTeX{{\rm B\kern-.05em{\sc i\kern-.025em b}\kern-.08em
    T\kern-.1667em\lower.7ex\hbox{E}\kern-.125emX}}

\makeatletter
\newcommand*\bigcdot{\mathpalette\bigcdot@{.5}}
\newcommand*\bigcdot@[2]{\mathbin{\vcenter{\hbox{\scalebox{#2}{$\m@th#1\bullet$}}}}}
\makeatother
\begin{document}

\title{MRNN: A Multi-Resolution Neural Network with Duplex Attention for Document Retrieval in the Context of Question Answering
}

\author{
\IEEEauthorblockN{Tolgahan Cakaloglu}
\IEEEauthorblockA{\textit{Computer Science} \\
\textit{University of Arkansas}\\
Little Rock, Arkansas \\
txcakaloglu@ualr.edu}
\and
\IEEEauthorblockN{Xiaowei Xu}
\IEEEauthorblockA{\textit{Information Science} \\
\textit{University of Arkansas}\\
Little Rock, Arkansas \\
xwxu@ualr.edu}
}

\maketitle
\IEEEpeerreviewmaketitle

\begin{abstract}
The primary goal of ad-hoc retrieval (document retrieval in the context of question answering) is to find relevant documents satisfied the information need posted in a natural language query. It requires a good understanding of the query and all the documents in a corpus, which is difficult because the meaning of natural language texts depends on the context, syntax, and semantics. Recently deep neural networks have been used to rank search results in response to a query. In this paper, we devise a \textit{multi-resolution neural network} (\textit{MRNN}) to leverage the whole hierarchy of representations for document retrieval. The proposed MRNN model is capable of deriving a representation that integrates representations of different levels of abstraction from all the layers of the learned hierarchical representation. Moreover, a duplex attention component is designed to refine the multi-resolution representation so that an optimal context for matching the query and document can be determined. More specifically, the first attention mechanism determines optimal context from the learned multi-resolution representation for the query and document. The latter attention mechanism aims to fine-tune the representation so that the query and the relevant document are closer in proximity. The empirical study shows that MRNN with the duplex attention is significantly superior to existing models used for ad-hoc retrieval on benchmark datasets including SQuAD, WikiQA, QUASAR, and TrecQA. 
\end{abstract}

\begin{IEEEkeywords}
Deep Learning, Ad-hoc Retrieval, Learning Representations, Ranking, Text Matching
\end{IEEEkeywords}

\section{Introduction}
Ad-hoc retrieval \cite{Voorhees2005}, document retrieval in the context of question answering, allows a user to specify the information need using a natural language query, which is instrumental for many applications including question answering and information retrieval. It requires a good understanding of the query and all the documents in a corpus, which is rather difficult because the meaning of natural language texts depends on the context, syntax, and semantics. Traditional approach uses simple statistical features such as term frequency and document frequency to represent the query and the document \cite{salton1986introduction}. The query and documents are matched by using some similarity measure like cosine similarity. However, this approach is less effective because the representation doesn't consider the rich context of texts. 

Recently deep neural networks have been used to rank search results in response to a query for ad-hoc retrieval \cite{Palangi2016DeepSE} \cite{McDonald2018DeepRR}. The fundamental idea of deep learning is that a hierarchical representation is learned automatically, where each layer is a representation that is a high-level abstraction of the representation from the previous layer. The most abstract representation from the last layer of the hierarchy is then used for the machine learning task. However, an abstract representation from a single layer or only a few layers may not be able to capture the semantic relationship of concepts across different levels of an ontology. A text may contain concepts from different levels of the ontology. For example, \textit{"a banana is a fruit"} where \textit{"fruit"} is a high level concept comparing to \textit{"banana"}.

\begin{figure}[h]
  \centering
  \includegraphics[width=0.50\textwidth]{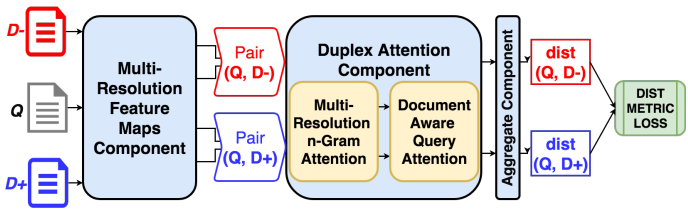}
  \caption{Overall training flow diagram for the proposed Multi-Resolution Neural Network with a query $Q$, a positive document $D^{+}$ and a negative document $D^{-}$}
  \label{fig:network_overview}
\end{figure}

In this paper, we present a new \textit{Multi-Resolution Neural Network} for ad-hoc retrieval, called \textit{MRNN}, which leverages representations across all levels of abstractions in the learned hierarchical representation. The learned representation achieves a multi-resolution effect that can represent concepts and their relationships across all the levels. As illustrated in Figure~\ref{fig:network_overview} the proposed model consists of the following components:
\begin{itemize}
    \item \textit{Multi-Resolution Feature Maps}, which transforms the input query and document into a multi-resolution representation.
    \item \textit{Duplex Attention}, which implements two attention mechanisms to refine the multi-resolution representation. The first attention mechanism determines an optimal context based on $n$-grams from the learned multi-resolution representation for the query and document. The latter attention mechanism aims to fine-tune the representation so that the query and the relevant document are closer in proximity. 
    \item \textit{Aggregation}, which calculates the similarity between a pair of query and document through aggregation of the duplex attention as the input to the loss function.
    \item \textit{Distance Metric Loss}, which is a loss function used to train the model by minimizing the loss.
\end{itemize}

The main contribution of the paper is as follows:
\begin{itemize}
    \item We first propose a new deep learning model MRNN that leverages representations across all levels of abstraction in the learned hierarchical representation for ad-hoc retrieval.
    \item A duplex attention mechanism is designed to refine the multi-resolution so that an optimal matching of the query and document can be achieved. More specifically, the first mechanism determines a proper context, and the latter fine-tunes the representation by considering the interplay between the query and document.
    \item The proposed model significantly outperforms existing models for ad-hoc retrieval on major benchmark datasets.
\end{itemize}

The rest of the paper is organized as follows. First, we review recent advances in ad-hoc retrieval in Section~\ref{relatedwork}. In Section~\ref{proposedapproach} we describe the details of the proposed model. An empirical study to compare the proposed method to the existing approaches is conducted. The experiment and the result are reported in Section~\ref{experiments} and Section~\ref{results} respectively. Finally, we conclude the paper with some future research in Section~\ref{conclusion}.

\section{Related Work} \label{relatedwork}
In the last few decades, machine learning methods have been practiced to information retrieval (IR) task, and show significant performances to this field.
% The researches towards to achieving the retrieval goal are generalized to the name of \textit{learning to rank} that is consists of three different aspects: \textit{pointwise}, \textit{pairwise}, and \textit{listwise}. These aspects have different parameters and goal to optimize. As it was well covered by previous works \cite{Liu2011LearningTR} \cite{Pang2017DeepRankAN}. The pointwise aspect \cite{Gey1994InferringPO} calculates a relevance score of each given document that is given to function as a feature vector. The pairwise aspects \cite{Freund1998AnEB}\cite{Joachims2002OptimizingSE} computes the optimal ordering for the given a pair of documents represented by feature vectors, and compare it to the ground truth. Listwise aspect \cite{Cao2007LearningTR}\cite{Burges2010FromRT}\cite{Xu2007AdaRankAB} identifies the ranked list based on given a set of document features associated with a query. These aspects need to learn useful features from the document and query vectors. Since the feature selection plays an important role to achieve the goal of the task, it leads that research to further improvement that automatically extracts better feature representations and also model architecture. 

Recently, deep learning techniques \cite{Lu2013ADA}\cite{Hu2014ConvolutionalNN}\cite{Palangi2016DeepSE}\cite{Guo2016ADR}\cite{Hui2017PACRRAP} have been evolved and applied to IR, in other words document relevance ranking also known as \textit{ad-hoc retrieval \cite{Voorhees2005}}. In the ad-hoc retrieval task, the number of words in documents are generally greater than the number of words in queries which prevent aforementioned methods from other tasks that focuses on pairs of short contents and some might not even be in natural language form, not suitable. Document ranking methods can be defined under the two titles: separation-oriented such as \cite{Palangi2016DeepSE} and interaction-oriented such as \cite{McDonald2018DeepRR}. In the separation-oriented, a query and a document representations are generated separately. At the final step, interactions of these documents are getting calculated through dot-product where a result shows the relevance. In the interaction-oriented, specific encodings between the query and document pairs are induced where it satisfies the exact matching as well as similarity matching that are the most important conditions for ad-hoc retrieval. 

In the area of machine reading-style question answering \cite{squad} \cite{Yang2015WikiQAAC} \cite{dhingra2017quasar} \cite{Wang2007WhatIT}, the system needs to find the answer in the given corresponding document or context. The models have to combine information retrieval and machine reading. Note that we do not benchmark the quality of the extraction phase, therefore we do not study extracting the answer from the retrieved document, but compare the quality of retrieval methods, and the feasibility of learning specialized neural models for retrieval purposes. DrQA \cite{Chen2017ReadingWT} is built on top of two component; a Document Retriever and a Document Reader respectively. The Document Retriever is a TF-IDF \cite{salton1986introduction} retrieval system
built upon Wikipedia corpus. Whereas, ConvRR \cite{tolgahan} is a convolutional residual retrieval network that focuses on achieving the retrieval performance using a hard triplet mining. WordCnt, WgtWordCnt, PV, PV+Cnt, and CNN+Cnt are the models derived from the following study \cite{Yang2015WikiQAAC}.Word Count method counts the number of non-stopwords in the question that also occur in the answer sentence, and Weighted Word Count re-weights the counts by the IDF values of the question words. PV represents the paragraph vector \cite{Le2014DistributedRO} where the result of PV is the similarity score between a question vector and document vector. CNN+Cnt is built on top of a bigram CNN \cite{Yu2014DeepLF} model with average pooling. PV+Cnt and CNN+Cnt are trained using a logistic regression classifier. QA-LSTM \cite{Tan2016ImprovedRL} is a biLSTM based model where the final representations of question and document are taken by max or mean pooling over all the hidden vectors. ABCNN \cite{Yin2016ABCNNAC} is an attention-based convolution neural network model that employs an attention feature matrix to influence convolutions to optimize the task. RNN-POA \cite{Chen2017EnhancingRN} is positional attention based RNN model that incorporates the positional content of the question words into the documents\textquotesingle \text{ }attentive representations. $SR^2$: Simple Ranker-Reader, $SR^3$: Reinforced Ranker-Reader \cite{Wang2018R3RR} are proposed to improve the performance of the machine reading-style question answering tasks. They utilized the Apache Lucene-based search engine and a deep neural network ranker that re-ranks the documents retrieved by the search engine incorporated by a machine reader. The latter model is trained using reinforcement learning. The results derived from the models show that the neural network ranker can learn to rank the documents based on semantic similarity with the question. InferSent Ranker \cite{Conneau2017SupervisedLO}\cite{Htut2018TrainingAR} is used to produce distributed representations for question and documents and, then, the input feature representation is built by concatenating the question representation, document representations, their difference, and their element-wise product. Receiving that input feature representation, the similarity score is calculated using a feed-forward neural network. Relation-Networks (RN) Ranker \cite{Santoro2017ASN}\cite{Htut2018TrainingAR}, further, targets on calculating the relevance or local interactions between words in the question and paragraph. Thus, this model is built to interpret the relation between question-document pairs. Tree Edit Model \cite{Heilman2010TreeEM} is represented as sequences of tree transformations involving complex reordering phenomena and demonstrate a  method for modeling pairs of semantically related contexts. They utilize a tree kernel in a greedy search routine to extract sequences of edits and use them in a logistic regression model to classify them. LSTM \cite{Wang2015ALS} uses a stacked bidirectional Long-Short Term Memory (BiLSTM) network to consecutively extract words from question and answer documents and then outputs their relevance scores. CNN \cite{Severyn2015LearningTR} is based on a convolutional neural network architecture for re-ranking pairs of short texts, where they teach the optimal representation of document pairs and a similarity function to relate them. AP-LSTM \cite{Tan2016ImprovedRL} is developed considering hybrid models that handle the documents using both convolutional and recurrent neural networks incorporating with attention mechanism to relate question and answer document.  AP-LSTM, AP-CNN \cite{Santos2016AttentivePN} are based on two-way attention mechanisms for discriminative model (CNN, RNN, LSTM) training.  AP allows the pooling layer to be aware of the input pair, in a way that information from the two can impact each other's representations. The model learns a similarity measure over projected $n$-grams of the pair, and generate the attention representation for each input to lead the pooling. Self-LSTM, Multihop-Sequential-LSTM \cite{Tran2018MultihopAN} are developed to expose the relations between question and answer document captured by attention. These models generate multiple representations that target on different parts of the question. Additionally, they utilize sequential attention mechanism which uses context information for computing context-aware attention weights. The proposed \textit{Multi-Resolution Neural Network} (MRNN) model belongs to the models of ad-hoc retrieval, which further incorporates the context of boosted interaction signals.

\section{Proposed Approach} \label{proposedapproach}
\subsection{Overview}
In this section, we introduce the proposed model that leverages the document relevance ranking called Multi-Resolution Neural Network as demonstrated in Figure~\ref{fig:network_overview} where the model is composed of two major components: While the initial component is responsible for generating $n$-gram feature maps, the latter one is matching those feature chains.

The model begins with a series of word inputs $e_1, e_2, e_3, ...., e_h$, that can establish a phrase, a sentence, a paragraph or a document. The model, then, builds \textit{Multi-Resolution Feature Maps} by densely connected $n$-Gram blocks. In order to conscientiously extract most useful features for the target retrieval and matching tasks, a \textit{Duplex Attention} component is introduced. The duplex attention component consists of two sub attention components that are called \textit{Multi-Resolution $n$-Gram Attention} and \textit{Document Aware Query Attention} respectively. The Multi-Resolution $n$-Gram Attention is responsible for reevaluating these Multi-Resolution feature maps while Document Aware Query Attention emphasizes matching those features between contextualized document and query to determine if they are similar or not. The final output is, then, used to improve the matching and retrieval performances.

\subsection{Multi-Resolution Feature Maps}
\begin{figure*}[h]
  \centering
  \includegraphics[width=0.99\textwidth]{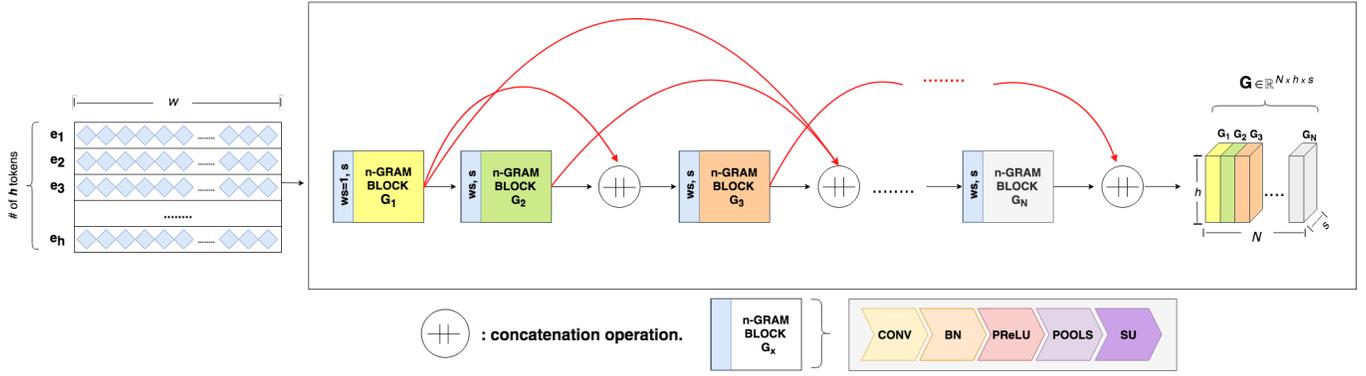}
  \caption{Multi-Resolution Feature Maps Component}
  \label{fig:feature_composition_component}
\end{figure*}

The first component called \textit{Multi-Resolution Feature Maps} of the proposed model is shown in Figure~\ref{fig:feature_composition_component}. The component begins with a series of word inputs that are initialized using one of the pre-trained embedding models. Let, $\textbf{e}_i \in \mathbb{R}^{w}$ be the $w$-dimensional embedding vector of $i$-th word of the input text. Then, a matrix that is composed of all words of the input text is represented as follows:
\begin{equation}
\textbf{\emE} = [\textbf{e}_1, \textbf{e}_2,\cdots, \textbf{e}_h]_{h\times w}
\end{equation}
where $\textbf{E}\in \mathbb{R}^{h\times w}$ be the $h\times w$-dimensional matrix of the input text and $h$ denotes the number of words in a given text. Embedded text, then, is fed to first $n$-Gram block of the component. 

\subsubsection{\textbf{$n$-Gram Blocks}}
A $n$-Gram block is representing five cascaded operations: a convolution (\textit{CONV}), a batch normalization (\textit{BN})\cite{Ioffe:2015}, a parametric rectified linear unit (\textit{PReLU})\cite{prelu}, pooling operations (\textit{POOLS}), and a scale unit \textit{SU}.
As the input text is represented with the matrix \textbf{\emE}, the output representations of each \textit{$n$-Gram} block within the component is denoted as below: 
\begin{equation}
\textbf{\emG}_{n} = [\textbf{g}_{n}^1, \textbf{g}_{n}^2,\cdots, \textbf{g}_{n}^h]_{h\times s}
\end{equation}
where $s$ denotes the dimension of the transformed feature representation, $n$ represents the index of an $n$-Gram block, and a total number of $n$-Gram blocks are labeled as $N$. Note that, while upstream $n$-Gram blocks are defined to emphasize the blocks that are located above the current block and close to the input layers of the network while the downstream blocks are located below the current block and close to the output layers of the network. In other words, feature maps derived from upstream blocks are associated with smaller $n$-grams (unigram, bigram, and etc), on the other hand, downstream blocks are associated with larger $n$-grams (6-grams, 7-grams, and etc) by reflecting on the feature maps derived from upstream $n$-Gram blocks.

Each $n$-Gram block utilizes $f_{gram}(\cdot,\cdot, \cdot)$ function to generate $\textbf{\emG}_{n}\in \mathbb{R}^{h\times s}$ representations. $f_{gram}(\cdot, \cdot, \cdot)$ can be described as the below: 
\begin{equation}
\textbf{\emG}_{n} = f_{gram}(\textbf{UB}, ws, s)
\end{equation}
where $\textbf{UB}$ denotes representations generated from upstream $n$-Gram blocks. $ws$, $s$ represents the window size and the dimension of the transformed feature representation respectively. The definition of $\textbf{UB}$ is conditioned to the index of the current $n$-Gram block.
\[
    \textbf{UB}= 
\begin{dcases}
    \textbf{\emE} ,& \text{if } n=1\\
    [\textbf{\emG}_{1},\textbf{\emG}_{2}, \cdots, \textbf{\emG}_{n-1}],& \text{otherwise}
\end{dcases}
\]
where $[\textbf{\emG}_{1},\textbf{\emG}_{2}, \cdots, \textbf{\emG}_{n-1}]$ shows the concatenation of the transformed representations derived from upstream $n$-Gram blocks ($1\leq n-1$), in other words, densely connected upstream $n$-Gram blocks.

\subsubsection{\textbf{Densely Connected Blocks}}
Conventional models that are designed by placing convolution blocks consecutively, aim to extract hierarchical feature representations. Specifically, for text-oriented task scenarios, convolutions can be evaluated as to derive $n$-gram features over a word sequence. Although traditional connections of convolution blocks extract hierarchical feature representations, they can not fulfill the requirements of natural languages for the following reasons:
\begin{itemize}
	\item Traditional convolution based networks exploit the kernels of a constant size where a constant size window slides across all text to generate feature representations \cite{Wang2017CombiningKW}. This is called constant size $n$-gram representations and it is not able to gather flexible size of $n$-gram representations that are needed for better understanding of the text that depends on context, syntax, and semantics.     
	\item In order to handle the extraction of flexible size $n$-gram representations, one can employ the kernels with various window sizes, but another issue is fired up with such settings: What would be the right architecture of using different kernel sizes? In other terms, how much expanding does the network require for different kernel sizes to produce the best features? It would end up with a huge search space for the greedy search due to an exponential number of parameter combinations. 
	\item Although the kernels with the variety of window sizes would still be seen as a better or an advanced architecture, it actually does not utilize the interplay between the representations derived from the different kernel size, since this type of approach consists of different independent parallel networks.
\end{itemize}
Therefore, we propose MRNN for ad-hoc retrieval, which leverages representations across all levels of abstractions in the learned hierarchical representation. The learned representation achieves a multi-resolution effect that can represent concepts and their relationship across all the levels of deep architecture. In order to consider all the resolutions (mixture of representations from adaptive $n$-grams such that feature maps of words or short phrases from the upstream blocks affect the downstream blocks to compose feature maps for longer context) of the representations throughout the proposed network, we first employ dense connections between each $n$-Gram blocks inspired by ideas from computer vision~\cite{densed} and text classifications ~\cite{Kim2018SemanticSM}~\cite{densetext}.

$f_{gram}(\cdot, \cdot, \cdot)$ has input parameters of upstream blocks representations $\textbf{UB}$, window size $ws$ and, dimension of the transformed representations $s$. Specifically, $f_{gram}(\cdot, \cdot, \cdot)$ computes the learnable weights $\textbf{W}_{n}\in \mathbb{R}^{s \times ws \times s}$ by using the cascaded aforementioned operations: \textit{CONV}, \textit{BN}, \textit{PReLU}, \textit{POOLS}, and \textit{SU}. The learnable weight tensor $\textbf{W}_{n}$ is composed of $s$ filters where each of them has a matrix $\in\mathbb{R}^{ws \times s}$, convolving $ws$ contiguous vectors. It is important to emphasize that in order to transform embedding dimension $w$ to $s$, in the case where $n=1$ or in other words $\textbf{UB}=\textbf{\emE}$, we employ $s$ filters where each of them has a matrix $\in\mathbb{R}^{(ws=1) \times w}$ in the first phase. Additionally to prevent the output size of the feature maps being different after each $n$-Gram block, we use padding and pooling operations. We, further, apply a scalar unit to the feature map. That scalar $sc$ is also learnable and it weights the feature map of the current $n$-th $n$-Gram block to decide how much the current block contributes to the downstream $n$-Gram blocks. Likewise, for the case where $\textbf{UB}=[\textbf{\emG}_{1},\textbf{\emG}_{2}, \cdots, \textbf{\emG}_{n-1}]$ or, in other words, where $n>1$, $f_{gram}(\cdot, \cdot, \cdot)$ computes the learnable weights $\textbf{W}_{n}\in \mathbb{R}^{(n-1) \times s \times ws \times s}$ by using the same cascaded operations in the block. After considering all $n$-Gram blocks, the multi-resolution feature maps tensor is represented as follows: $\textbf{\emG} = [\textbf{\emG}_{1},\textbf{\emG}_{2}, \cdots, \textbf{\emG}_{N}] \in\mathbb{R}^{N \times h \times s}$ and $\textbf{\emG}$ consists of $h$ matrices where each of them is representing the feature map matrix $\textbf{\emG}_{n}$ from each $n$-Gram block.

\subsection{Duplex Attention Component}
Duplex attention component is one of the most important components of our proposed network. Since attention~\cite{attention_} evolves into an effective component within the neural network for extracting useful information, which achieves a remarkable result for many machine learning tasks, we compose this component with two sub attention components that are called \textit{Multi-Resolution $n$-Gram Attention} and \textit{Document Aware Query Attention} respectively. The multi-resolution feature maps tensors of each query($q$)-document($d$) pair are denoted as $\textbf{Gq}$ - $\textbf{Gd}$. Hence, the Multi-Resolution $n$-Gram Attention is responsible for reevaluating $\textbf{Gq}$, and $\textbf{Gd}$ feature maps while Document Aware Query Attention emphasizes matching those features between contextualized document and query to determine if they are similar or not. 

\subsubsection{\textbf{Multi-Resolution $n$-Gram Attention}}

\begin{figure*}[h]
  \centering
  \includegraphics[width=0.99\textwidth]{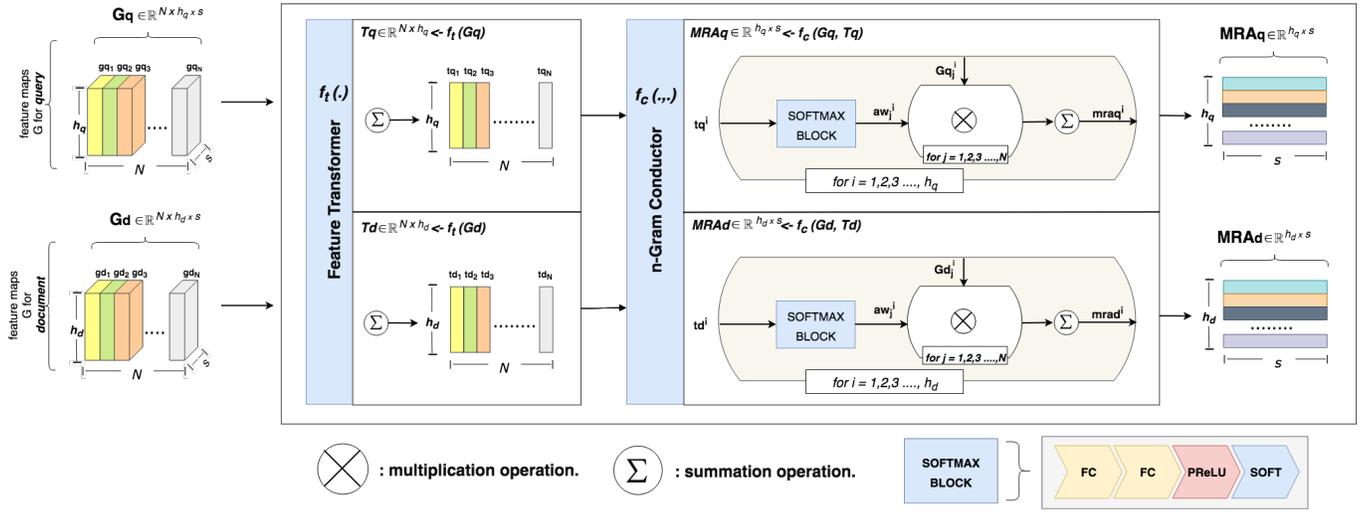}
  \caption{Multi-Resolution $n$-Gram Attention}
  \label{fig:multi_resolution_att}
\end{figure*}

The multi-resolution feature maps tensor $\textbf{\emG}$ ($\textbf{Gq}$ or $\textbf{Gd}$) contains feature maps from all $n$-Gram blocks. More specifically, $\textbf{Gq} = [\textbf{Gq}_{1},\textbf{Gq}_{2}, \cdots, \textbf{Gq}_{N}] \in\mathbb{R}^{N \times h_{q} \times s}$, and $\textbf{Gd} = [\textbf{Gd}_{1},\textbf{Gd}_{2}, \cdots, \textbf{Gd}_{N}] \in\mathbb{R}^{N \times h_{d} \times s}$, the number of words in the query and the document are denoted as $h_{q}$ and $h_{d}$ respectively. Although the network has rich features at this step, some of those features still need to be pruned. In order to prune them conscientiously for the next component, we introduce a multi-resolution $n$-Gram attention component as shown in Figure~\ref{fig:multi_resolution_att}.  The multi-resolution $n$-Gram attention component has two consecutive functions called transformer  $f_{t}(\cdot)$ and conductor $f_{c}(\cdot, \cdot)$ respectively. For the sake of simplicity, we explain the functions by taking the multi-resolution feature maps of the query ($\textbf{Gq}$) into consideration. 

The transformer $f_{t}(\cdot)$ is a function that is formalized as below:
\begin{equation}
\textbf{Tq} = f_{t}(\textbf{Gq})
\end{equation}
where $\textbf{Tq}\in\mathbb{R}^{N \times h_{q}}$ is a matrix of scalar adjusters. Since we know that  $\textbf{Gq}_n = [\textbf{gq}_{n}^1,\textbf{gq}_{n}^2, \cdots,\textbf{gq}_{n}^{h_q}]_{h_q \times s}$ and each $\textbf{gq}_{n}^i$ represents the $s$ dimensional feature representation in the $i$-th location of a query at $n$-th $n$-Gram block. Particularly, for each $\textbf{gq}_{n}^i$ feature representation, $f_{t}(\cdot)$ calculates the the scalar adjuster vector $\textbf{tq}^{i}$:
\begin{equation}
\textbf{tq}^{i} = [\sum_{j=1}^{s}\textbf{gq}_{1}^i[j], \sum_{j=1}^{s}\textbf{gq}_{2}^i[j], \cdots, \sum_{j=1}^{s}\textbf{gq}_{n}^i[j]]
\end{equation}
where $\textbf{tq}^i\in\mathbb{R}^{N}$ is a $N$-dimensional vector, thus,  $\textbf{Tq}=[\textbf{tq}^1, \textbf{tq}^2, \cdots, \\ \textbf{tq}^{h_q}]$ is a matrix of scalar adjusters. The motivation behind this procedure is that the sum of all the values in the $s$-dimensional vector of $\textbf{gq}_{n}^i$ is positioned as feature importance. 

The scalar adjusters matrix $\textbf{Tq}$ and the multi-resolution feature map tensor of query $\textbf{Gq}$ are, further, passed to conductor function $f_{c}(\cdot, \cdot)$ to conduct the feature maps from variety of $n$-gram scales by calculating attention weights via the softmax block. $f_{c}(\cdot, \cdot)$ is described:
\begin{equation}
\textbf{MRAq} = f_{c}(\textbf{Gq}, \textbf{Tq})
\end{equation}
where $\textbf{MRAq}\in\mathbb{R}^{h_{q} \times s}$ is a matrix of multi-resolution $n$ gram attention vectors. For each scalar vector $\textbf{tq}^{i}$, $f_{c}(\cdot, \cdot)$, first, calculates the attention weights via softmax block represented as: 
\begin{equation}
\textbf{aw}^{i}= f(\textbf{tq}^i)
\end{equation}
where $f(\cdot)$ defines the softmax block that is a perceptron of the following cascaded operations: fully connected layers (\textit{FC}), a parametric rectified linear unit (\textit{PReLU}), and a softmax operation (\textit{SOFT}). $\textbf{aw}\in\mathbb{R}^{N}$ is an attention weights vector. Thus, representations of $\textbf{tq}^i$ and $\textbf{aw}^i$  can be shown as:  
\[ 
\textbf{tq}^i = [{tq}^i_1, {tq}^i_2, \cdots, {tq}^i_N]
\]
\[ 
\textbf{aw}^i = [{aw}^i_1, {aw}^i_2, \cdots, {aw}^i_N]
\]
As next steps,  $f_{c}(\cdot, \cdot)$ computes the final multi-resolution $n$ gram attention vector using the following equations: 
\begin{equation}
\textbf{mraq}^i = \sum_{j=1}^{N}\textbf{aw}^{i}_{j} \cdot \textbf{Gq}_{j}^i 
\end{equation}
where $\textbf{mraq}^i \in\mathbb{R}^{s}$ is a multi-resolution $n$ gram attention vector of $i$-th position of a query. The final output representations of $f_{c}(\cdot, \cdot)$ is a multi-resolution $n$ gram attention matrix \textbf{MRAq} that is evaluated as:
\[ 
\textbf{MRAq} = [\textbf{mraq}^1,\textbf{mraq}^2, \cdots, \textbf{mraq}^{h_q}] \in\mathbb{R}^{h_{q} \times s}
\]
Likewise, $\textbf{MRAd}\in\mathbb{R}^{h_{d} \times s}$  is the multi-resolution $n$-Gram attention matrices for document. Both of matrices, then, are passed to next attention component called \textit{Document Aware Query Attention}.

\subsubsection{\textbf{Document Aware Query Attention}}
\begin{figure*}[h]
  \centering
  \includegraphics[width=0.99\textwidth]{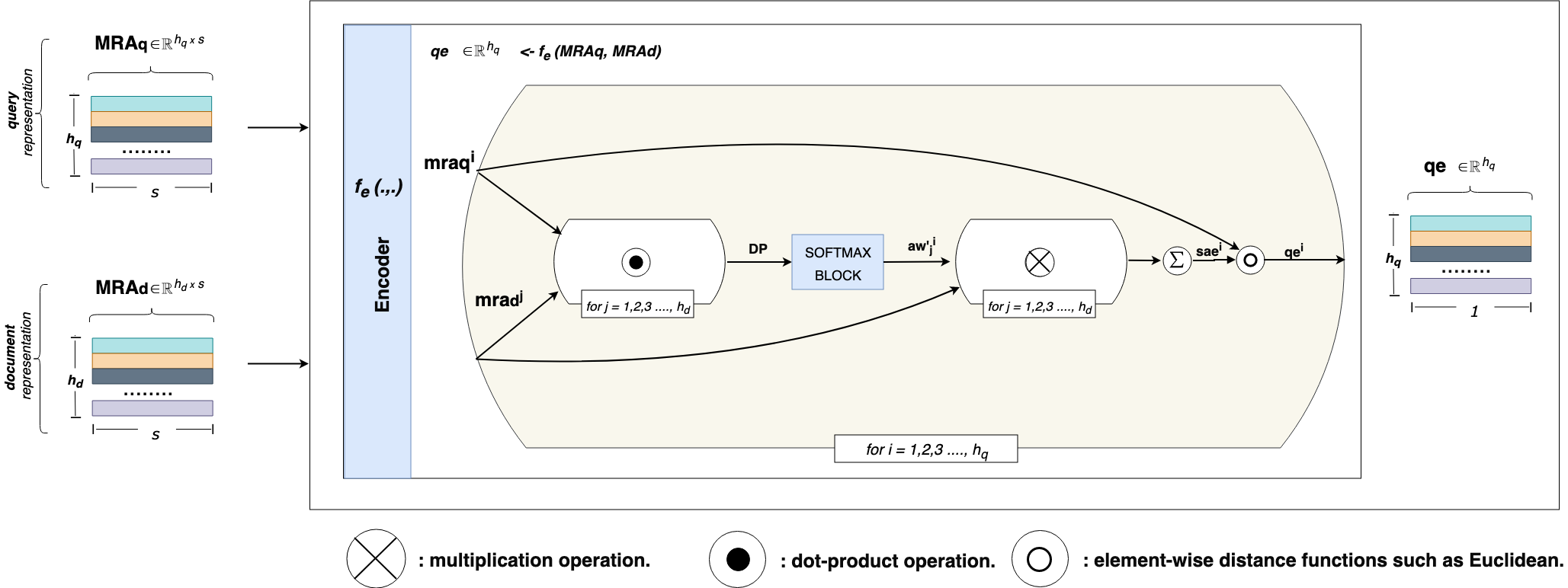}
  \caption{Document Aware Query Attention}
  \label{fig:document_aware_att}
\end{figure*}

In order to compute document aware query encoding using attention mechanism, we present \textit{Document Aware Query Attention} component as illustrated in Figure~\ref{fig:document_aware_att}. The component emphasizes matching features between the multi-resolution $n$-Gram attention matrices of query $\textbf{MRAq}\in\mathbb{R}^{h_{q} \times s}$ and document $\textbf{MRAd}\in\mathbb{R}^{h_{d} \times s}$ pair. The document aware query attention component has a function called encoder $f_{e}(\cdot, \cdot)$. 

$f_{e}(\cdot, \cdot)$ is an encoder function that is formalized as below:
\begin{equation}
\textbf{qe} = f_{e}(\textbf{MRAq}, \textbf{MRAd})
\end{equation}
where $\textbf{qe}\in\mathbb{R}^{h_{q}}$ is a vector of document aware query encodings via attention weights. We, first, calculate a dot-product attention weights $aw^{\prime}_j$ for each position of $\textbf{MRAd}$ relative to $\textbf{mraq}^{i}$ via the softmax block as follows: 
\begin{equation}
\textbf{aw}^{\prime i} = f(\textbf{mraq}^{i} \bigcdot \textbf{MRAd})
\end{equation}
where $f(\cdot)$ defines the softmax block that is a perceptron with the same architecture aforementioned (\textit{FCs}, \textit{PReLU}, \textit{SOFT}). $\textbf{aw}^{\prime i}\in\mathbb{R}^{h_{d}}$ is an attention weights vector. Thus, representation $\textbf{aw}^{\prime i}$  can be shown as:  
\[ 
\textbf{aw}^{\prime i} = [{aw}^{\prime i}_1, {aw}^{\prime i}_2, \cdots, {aw}^{\prime i}_{h_d}]
\]
As a next step, we sum the document aware encodings of the $h_d$-locations, scaled by their attention weights, to create an attention-based representation $\textbf{sae}^{i}$ of document representations \textbf{MRAd} from the aspect of query representation $\textbf{mraq}_{i}$ formulated as below:
\begin{equation}
\textbf{sae}^i = \sum_{j=1}^{h_{d}}\textbf{aw}^{\prime i}_{j} \cdot \textbf{mrad}^{j} 
\end{equation}
The element-wise distance (euclidean) between the attention-based document representation $\textbf{sae}^{i}$ and $\textbf{mraq}^{i}$ is, further, calculated and adopted as document aware query encoding $qe^i \in\mathbb{R}^{h_q}$ that can be defined as a document aware query encoding of $i$-th position of a query. The final output representations of encoder $f_{e}(\cdot, \cdot)$ is a document aware query encoding attention vector \textbf{qe} that is evaluated as:
\[ 
\textbf{qe} = [qe^1,qe^2, \cdots, qe^{h_q}] \in\mathbb{R}^{h_{q}}
\]
In other words, if the document includes more positions of $\textbf{mrad}^{j}$ that are very much alike to the position of $\textbf{mraq}^{i}$ in the query, the document aware query attention component indicates mostly those positions and, therefore $\textbf{sae}^i$ will be close to $\textbf{mraq}^{i}$. 

As a final step, we aggregate these similarities $qe^i$ via the \textit{aggregate component} to define the distance of query-document pair as follows:
\begin{equation}
dist = \sum_{i=1}^{h_{q}}\textbf{qe}^{i}
\end{equation}
where $dist\in\mathbb{R}$ is the final distance descriptor between the query-document pair. 

\subsection{Distance Metric Loss Function}
In order to train the proposed MRNN to perform well on retrieval and matching tasks, which also generalizes well on unseen data, we utilized \textit{triplet loss}\cite{Schroff2015FaceNetAU} during the training period as shown in Figure~\ref{fig:network_overview}. With this setup, the network is encouraged to reduce distances between positive pairs so that they are smaller than negative ones. A particular query $Q$ would be a query anchor close in proximity to a document $D^+$ as the positive pair to the same question than to any document $D^-$ as they are positive pairs to other questions. The key point of the $L_\text{triplet}$ is to build the correct triplet structure which should meet the condition of the following equation:
\[
\rVert Q, D^+ \rVert^{2} + m < \rVert Q, D^- \rVert^{2}
\]
For each query, the document $D^+$ is selected as:
$\argmax_{D^+}\rVert Q, D^+ \rVert^{2}$ and likewise the hardest document $D^-$:
$\argmin_{D^-}\rVert Q, D^- \rVert^{2}$ to form a triplet. This triplet selection strategy is called \textit{hard triplets mining}. 

Let $\textbf{T} = (Q,D^+,D^-)$ be a triplet input. Given $\textbf{T}$, the proposed approach computes the distances between the positive and negative pairs via the proposed MRNN.
\begin{equation}
L_\text{triplet} = [\rVert Q, D^+ \rVert^{2} - \rVert Q, D^- \rVert^{2}+ m]^+
\end{equation}
where $m > 0$ is a scalar value called margin, and $\rVert \cdot, \cdot  \rVert^2$ represents the distance score between two objects.
\section{Experiments} \label{experiments}
\subsection{Datasets}
In order to evaluate our proposed approach, we conducted extensive experiments on four datasets, including SQuAD \cite{squad}, WikiQA \cite{Yang2015WikiQAAC}, QUASAR \cite{dhingra2017quasar}, and TrecQA \cite{Wang2007WhatIT}.

\subsubsection{\textbf{SQuAD}}
The Stanford Question Answering Dataset (SQuAD) \cite{squad} is a large reading comprehension dataset that is built with $100,000+$ questions. Each of these questions is composed by crowd workers on a set of Wikipedia documents where the answer to each question is a segment of text from the corresponding reading passage. In other words, the consolidation of retrieval and extraction tasks are aimed at measuring the success of the proposed systems.

\subsubsection{\textbf{WikiQA}}
The Wikipedia open-domain Question Answering (WikiQA) \cite{Yang2015WikiQAAC} dataset is collected using Bing query logs. For each question, clicked Wikipedia pages (issued by at least 5 unique users) and used sentences in the summary section of Wikipedia page as the candidates, is further marked on a crowdsourcing platform. Note that we excluded questions that have no candidates. Based on training, dev and test subsets,  $1,242$ questions are used in the experiment. 

\subsubsection{\textbf{QUASAR}}
The Question Answering by Search And Reading (QUASAR) is a large-scale dataset consisting of QUASAR-S and QUASAR-T. Each of these datasets is built to focus on evaluating systems devised to understand a natural language query, a large corpus of texts and to extract an answer to the question from the corpus. Similar to SQuAD QUASAR is primarily used to measure the success of the proposed systems for ad-hoc retrieval and extraction tasks. Specifically, QUASAR-S comprises $37,012$ fill-in-the-gaps questions that are collected from the popular website Stack Overflow using entity tags. Since our research is not to address the fill-in-the-gaps questions, we want to pay attention to the QUASAR-T dataset that fulfills the requirements of our focused retrieval task. The QUASAR-T dataset contains $43,012$ open-domain questions collected from various internet sources. The candidate documents for each question in this dataset are retrieved from an Apache Lucene based search engine built on top of the ClueWeb09 dataset \cite{callan2009clueweb09}.

\subsubsection{\textbf{TrecQA}}
Text Retrieval Conference Question Answering (TrecQA) is a popular benchmark dataset for question answering prepared by \cite{Wang2007WhatIT}. TrecQA dataset is based on QA track (8-13) of TREC. The dataset consists of factoid questions, each of which has a single sentence as a candidate answer. In order to make the comparison parallel to the previous works, we pursue the same strategy they applied where all questions with only positive or negative answers are excluded. In total, we end up having $1,295$ questions within all training, dev and test subsets of TrecQA. 

The number of queries in each dataset including their subsets is listed in Table~\ref{dataset_statistic}. 
\begin{table}[ht]
\caption{Datasets Statistics: Number of queries in each train, validation, and test subsets}
\label{dataset_statistic}
\vskip 0.001in
\begin{center}
\begin{small}
\begin{sc}
\begin{tabular}{lccccr}
\toprule
Dataset & train & valid. & test & total \\
\midrule
SQuAD & 87,599 & 10,570 & hidden& 98,169+\\
WikiQA & 873 & 126 & 243& 1,242\\
QUASAR-T & 37,012& 3,000& 3,000& 43,012\\
TrecQA & 1,162& 65& 68& 1,295\\
\bottomrule
\end{tabular}
\end{sc}
\end{small}
\end{center}
\vskip -0.1in
\end{table}

\subsection{Performance Measure}
The matching and retrieval tasks aim to improve the $recall@k$, the Mean Reciprocal Rank ($MRR$) and Mean Average Precision ($MAP$). The $recall@k$ score is calculated by selecting the correct pair among all candidates. Basically, $recall@k$ is defined as the number of correct documents as listed within top-$k$ returns out all possible documents. Likewise, $MRR$ and $MAP$ metrics are also commonly used in information retrieval and question answering researches \cite{Manning:2008:IIR:1394399}.
% Additionally, embedding representations are visualized using t-distributed stochastic neighbor embedding by \citet{Maaten2008VisualizingDU} in order to project the clustered distributions of the questions that are assigned to same documents.
\subsection{Implementation}
\subsubsection{\textbf{Input}}
We adopt the multi-resolution word embedding \cite{tolgahan} using Bert\cite{devlin2018bert}, ELMo\cite{elmo}, FastText\cite{mikolov2018advances} for each question and document in datasets. We configure the multi-resolution word embedding as authors stated in their work:  $f_{mixture}(\cdot,\cdot,\cdot, \cdot)$ and $f_{ensemble}(\cdot,\cdot)$ configurations are shown in Table~\ref{multi_resolution_mixture} and Table~\ref{multi_resolution_ensemble} respectively.

\begin{table}[ht]
\caption{$f_{mixture}(\cdot,\cdot,\cdot)$ configuration of the multi-resolution word embedding}
\label{multi_resolution_mixture}
\vskip 0.001in
\begin{center}
\begin{small}
\begin{sc}
\begin{tabular}{lcccr}
\toprule
%function name & \textbf{E} & w_{idf} & \textbf{m} & \gamma \\
\textbf{E} & $w_{idf}$ & \textbf{$m$} & $f_{mix}$ & out \\
\midrule
Bert & False & [$\frac{1}{4}$, $\frac{1}{4}$, $\frac{1}{4}$, $\frac{1}{4}$,0,..,0] & $concat.$ &$\textbf{x}^1$\\
ELMo & True & [0, 0, 1] & $sum$ &$\textbf{x}^2$\\
FastText & True & [1] & $sum$  &$\textbf{x}^3$\\
\bottomrule
\end{tabular}
\end{sc}
\end{small}
\end{center}
\vskip -0.1in
\end{table}

\begin{table}[ht]
\caption{$f_{ensemble}(\cdot)$ configuration of the multi-resolution word embedding}
\label{multi_resolution_ensemble}
\vskip 0.001in
\begin{center}
\begin{small}
\begin{sc}
\begin{tabular}{lcr}
\toprule
\textbf{X'} & \textbf{$u$} & $f_{ensemble}$ \\
\midrule
\{$\textbf{x}^1$, $\textbf{x}^2$, $\textbf{x}^3$\}& [$\frac{1}{3}$, $\frac{1}{3}$, $\frac{1}{3}$] & $concat.$\\
\bottomrule
\end{tabular}
\end{sc}
\end{small}
\end{center}
\vskip -0.1in
\end{table}

\subsubsection{\textbf{MRNN Training Configuration}}
The proposed MRNN is implemented with Tensorflow 1.8+ by \cite{tensorflow2015-whitepaper} and trained on NVIDIA Tesla K40c GPUs. Specifically, the network is trained using ADAM optimizer \cite{adam} with a batch size of $512$. The learning rate is set to $10^{-4}$. Additionally, the weight decay is set to $10^{-3}$ to tackle over-fitting. The triplet loss is, then, chosen as an objective function with different margins for each of the datasets. ($m=1$: SQuAD, $m=0.8$: QUASAR-T, $m=0.5$: \{WikiQA ,TrecQA\}). We configured 6 $n$-Gram blocks ($N=6$) for SQuAD and QUASAR-T datasets, and 4 $n$-Gram blocks ($N=4$) for WikiQA and TrecQA datasets. Last but not least, window size and transformed feature representation dimension are set to $ws = 3$, $s=1024$ respectively.

\section{Results}\label{results}
We compare our approach with different models proposed by other researchers for each dataset using their evaluation measures and test subsets. 
\subsection{\textbf{SQuAD}}
The $recall@5$ result is calculated for SQuAD in order to compare with the document retrieval component of the multi-layer recurrent neural network \cite{Chen2017ReadingWT} (DrQA) and the convolutional residual retrieval network (ConvRR) \cite{tolgahan}\cite{tolgahan1}. The comparisons are shown in Table~\ref{table-squad}.

\subsection{\textbf{WikiQA}}
WordCnt, WgtWordCnt, and CNN-Cnt are the models derived from the following initial study \cite{Yang2015WikiQAAC}. On top of the baseline models, the Paragraph Vector (PV) and PV + Cnt models \cite{Yih2013QuestionAU} are taken into consideration. We, further, consider even more advanced models: QA-LSTM \cite{Tan2016ImprovedRL}, Self-LSTM, Multihop-Sequential-LSTM \cite{Tran2018MultihopAN}, ABCNN \cite{Yin2016ABCNNAC}, Rank MP-CNN \cite{Rao2016NoiseContrastiveEF}, RNN-POA \cite{Chen2017EnhancingRN}. In order to compare the proposed model with the aforementioned models for the WikiQA dataset, we calculate the Mean Reciprocal Rank ($MRR$) and Mean Average Precision ($MAP$) metrics and all the results are presented in Table~\ref{table-wikiqa}.

\subsection{\textbf{QUASAR}}
BM25 \cite{bm25}\cite{Htut2018TrainingAR}, $SR^2$: Simple Ranker-Reader, $SR^3$: Reinforced Ranker-Reader \cite{Wang2018R3RR}, InferSent Ranker \cite{Conneau2017SupervisedLO}\cite{Htut2018TrainingAR}, convolutional residual retrieval network (ConvRR) \cite{tolgahan}, and Relation-Networks (RN) Ranker \cite{Santoro2017ASN}\cite{Htut2018TrainingAR} are the models that are evaluated using the $recall@1$, $recall@3$, and $recall@5$. The comparisons are listed in Table~\ref{table-quasar}.

\subsection{\textbf{TrecQA}}
We compute the Mean Reciprocal Rank ($MRR$) and Mean Average Precision ($MAP$) metrics for the proposed model as well as following models: 
Tree Edit Model \cite{Heilman2010TreeEM}, LSTM \cite{Wang2015ALS}, CNN \cite{Severyn2015LearningTR}, AP-LSTM \cite{Tan2016ImprovedRL}, AP-CNN \cite{Santos2016AttentivePN}, RNN-POA \cite{Chen2017EnhancingRN}, and  Self-LSTM, Multihop-Sequential-LSTM \cite{Tran2018MultihopAN}. The results are stated in Table~\ref{table-trecqa}.

\subsection{Evaluation}
The result on all benchmark datasets shows that the proposed MRNN clearly outperforms existing models. Relation-Networks Ranker is the only model achieved a slightly better result for recall@3 and recall@5 on QUASAR-T dataset, where the proposed MRNN model still wins the most important recall@1.  
\begin{table}[ht]
\caption{Performances on SQUAD. $recall@k$ retrieved documents using the baseline models and the proposed model.}
\label{table-squad}
\vskip 0.15in
\begin{center}
\begin{small}
\begin{sc}
\begin{tabular}{lccr}
\toprule
Model & @5 \\
\midrule
ConvRR & 75.6\\
DrQA document-retrieval & 77.8\\
\midrule
MRNN & \textbf{80.4}\\
\bottomrule
\end{tabular}
\end{sc}
\end{small}
\end{center}
\vskip -0.1in
\end{table}

\begin{table}[ht]
\caption{Performances on WikiQA. The baseline models and the proposed model are listed based on the results derived from $MAP$ and $MRR$ metrics.}
\label{table-wikiqa}
\vskip 0.15in
\begin{center}
\begin{small}
\begin{sc}
\begin{tabular}{lcccr}
\toprule
Model & MAP & MRR \\
\midrule
WordCount & 0.4891 & 0.4924 \\
WgtWordCnt & 0.5099 & 0.5132 \\
PV & 0.511 & 0.516 \\
PV + Cnt & 0.599 & 0.609 \\
CNN + Cnt & 0.652	& 0.6652 \\
QA-LSTM & 0.654 & 0.665 \\
AP-LSTM & 0.670 & 0.684 \\
AP-CNN & 0.689 & 0.696 \\
Self-LSTM & 0.693 & 0.704 \\
ABCNN & 0.692 & 0.71 \\
Rank MP-CNN & 0.701 & 0.718 \\
RNN-POA & 0.721 & 0.731 \\
Multihop-Sequential-LSTM & 0.722 & 0.738 \\
\midrule
MRNN & \textbf{0.731} & \textbf{0.745} \\
\bottomrule
\end{tabular}
\end{sc}
\end{small}
\end{center}
\vskip -0.1in
\end{table}

\begin{table}[ht]
\caption{Performances on QUASAR-T. $recall@k$ retrieved documents using the baseline models and the proposed model.}
\label{table-quasar}
\vskip 0.15in
\begin{center}
\begin{small}
\begin{sc}
\begin{tabular}{lccccr}
\toprule
Model & @1 & @3 & @5 \\
\midrule
BM25 & 19.7 & 36.3 & 44.3 \\
$SR^2$: Simple Ranker-Reader & 28.8 & 46.4 & 54.9 \\
InferSent Ranker & 36.1 & 52.8 & 56.7 \\
$SR^3$: Reinforced Ranker-Reader & 40.3 & 51.3 & 54.5 \\  ConvRR & 50.67 & 63.1 & 67.4 \\ 
Relation-Networks Ranker & 51.4 & \textbf{68.2} & \textbf{70.3}\\
\midrule
MRNN & \textbf{52.8} & 67.7 & 69.9\\
\bottomrule
\end{tabular}
\end{sc}
\end{small}
\end{center}
\vskip -0.1in
\end{table}

\begin{table}[ht]
\caption{Performances on TrecQA. The baseline models and the proposed model are listed based on the results derived from $MAP$ and $MRR$ metrics.}
\label{table-trecqa}
\vskip 0.15in
\begin{center}
\begin{small}
\begin{sc}
\begin{tabular}{lcccr}
\toprule
Model & MAP & MRR \\
\midrule
Tree Edit Model & 0.609 & 0.692 \\ 
LSTM & 0.713 & 0.791 \\
CNN & 0.746 & 0.808 \\
AP-LSTM & 0.753 & 0.830 \\
AP-CNN & 0.753 & 0.851 \\
Self-LSTM & 0.759 & 0.830 \\ 
RNN-POA & 0.781 & 0.851 \\
Multihop-Sequential-LSTM & 0.813 & 0.893 \\
\midrule
MRNN & \textbf{0.822} & \textbf{0.898}\\
\bottomrule
\end{tabular}
\end{sc}
\end{small}
\end{center}
\vskip -0.1in
\end{table}

\subsection{Visualization and Analysis}
\begin{figure*}[h]
  \centering
  \includegraphics[width=0.99\textwidth]{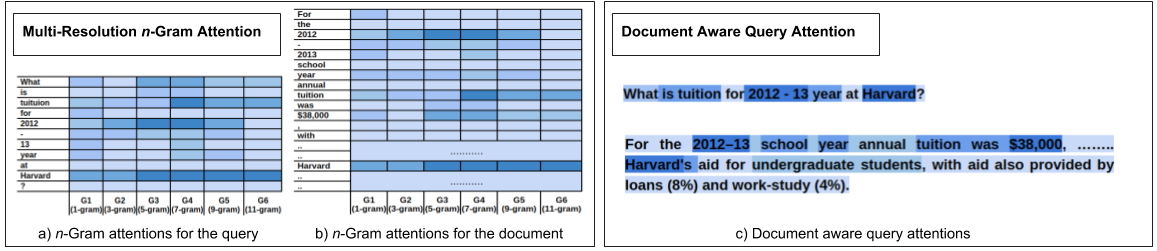}
  \caption{For the sample query and document extracted from SQuAD, the visualization of their attention weights that are fired in the sub attention components of Duplex Attention}
  \label{fig:heatmap}
\end{figure*}

The attention visualizations of the duplex attention component for the sample query and corresponding document extracted from the SQuAD dataset are shown in Figure~\ref{fig:heatmap}. When window-size is set to $3$ ($ws=3$), and the number of $n$-Gram blocks is set to $6$ ($N=6$), then each row of Figure~\ref{fig:heatmap},  a) and b) indicate an attention weight distribution over $\textbf{gq}_{n}^i$ as well as  $\textbf{gd}_{n}^i$ and $i$ is the position of each words in the matrices of the query and document. Hence, $\textbf{Gq}_{n}$ and $\textbf{Gd}_{n}$ corresponds to feature maps of (\textit{2n-1})-gram, e.g., $\textbf{Gq}_{1}$: 1-gram, $\textbf{Gq}_{2}$:3-gram, ... etc. The proposed MRNN puts more emphasizes on some segments of the query and the parts of the document as in the multi-resolution $n$-Gram attention. In the document aware query attention, MRNN gives more attention to additional segments of the question and also some
other parts of the document that have crucial interactions, as shown in Figure~\ref{fig:heatmap}, c).

\section{Conclusion}\label{conclusion}
Ad-hoc retrieval is an important task for question answering and information retrieval. This paper proposes a new multi-resolution neural network for ad-hoc retrieval, which is the first model that leverage the strength of representations of different abstract levels in the learned hierarchical representation. The proposed model incorporated with a new duplex attention mechanism can significantly improve the performance of ad-hoc retrieval. The experiment shows a superior result in comparison with other existing methods on some major benchmark datasets. In the future, we want to apply the proposed model to other areas including pattern recognition and computer vision.   

\bibliographystyle{IEEEtran}
\bibliography{sample-base}

% \begin{thebibliography}{00}

% \bibitem{b1} G. Eason, B. Noble, and I. N. Sneddon, ``On certain integrals of Lipschitz-Hankel type involving products of Bessel functions,'' Phil. Trans. Roy. Soc. London, vol. A247, pp. 529--551, April 1955.
% \bibitem{b2} J. Clerk Maxwell, A Treatise on Electricity and Magnetism, 3rd ed., vol. 2. Oxford: Clarendon, 1892, pp.68--73.
% \bibitem{b3} I. S. Jacobs and C. P. Bean, ``Fine particles, thin films and exchange anisotropy,'' in Magnetism, vol. III, G. T. Rado and H. Suhl, Eds. New York: Academic, 1963, pp. 271--350.
% \bibitem{b4} K. Elissa, ``Title of paper if known,'' unpublished.
% \bibitem{b5} R. Nicole, ``Title of paper with only first word capitalized,'' J. Name Stand. Abbrev., in press.
% \bibitem{b6} Y. Yorozu, M. Hirano, K. Oka, and Y. Tagawa, ``Electron spectroscopy studies on magneto-optical media and plastic substrate interface,'' IEEE Transl. J. Magn. Japan, vol. 2, pp. 740--741, August 1987 [Digests 9th Annual Conf. Magnetics Japan, p. 301, 1982].
% \bibitem{b7} M. Young, The Technical Writer's Handbook. Mill Valley, CA: University Science, 1989.
% \end{thebibliography}
% \vspace{12pt}
% \color{red}
% IEEE conference templates contain guidance text for composing and formatting conference papers. Please ensure that all template text is removed from your conference paper prior to submission to the conference. Failure to remove the template text from your paper may result in your paper not being published.

\end{document}